# Simulation de trajectoires de processus continus

Frédéric Planchet[1] et **Pierre-E. Thérond**[2]

**Résumé.** Les processus stochastiques continus sont des outils largement employés en finance et en assurance, notamment pour modéliser taux d'intérêts et cours d'actions. De nombreuses problématiques amènent à la simulation des trajectoires de tels processus. La mise en œuvre pratique de ces simulations nécessite de discrétiser ces processus, d'estimer les paramètres et de générer des nombres aléatoires en vue de générer les trajectoires voulues. Nous abordons successivement ces trois points que nous illustrons dans quelques situations simples.

**Mots-clés :** Simulation, discrétisation de processus continus, schémas d'Euler et de Milstein, estimation par inférence indirecte, générateur du *tore mélangé*.

**Abstract.** Continuous time stochastic processes are useful models especially for financial and insurance purposes. The numerical simulation of such models is dependant of the time discrete discretization, of the parametric estimation and of the choice of a random number generator. The aim of this paper is to provide the tools for the practical implementation of diffusion processes simulation, particularly for insurance contexts.

## 1 Introduction

La fréquence des variations de cotation des actifs financiers sur les marchés organisés a conduit les financiers à considérer des processus stochastiques continus pour modéliser les évolutions de taux d'intérêt comme de cours d'actions. Si les outils mathématiques sous-jacents peuvent, de prime abord, sembler plus complexes que leurs équivalents discrets, leur usage a notamment permis d'obtenir des formules explicites pour l'évaluation d'actifs contingents (*cf.* les formules fermées de prix d'options européennes obtenues par BLACK et SCHOLES [1973] lorsque l'actif sous-jacent suit un mouvement brownien géométrique). Ces modèles sont aujourd'hui utilisés dans de nombreux domaines, notamment en assurance. Leur mise en œuvre pratique nécessite de les discrétiser, que ce soit pour l'estimation des paramètres ou pour la simulation des trajectoires.

En effet les données disponibles étant discrètes, l'estimation des paramètres des modèles en temps continu n'est pas immédiate. De nombreux processus, tels que le modèle de taux proposé par COX, INGERSOLL et ROSS [1985], n'admettant pas de discrétisation exacte, l'estimation des paramètres nécessite souvent une approximation discrétisée. Les estimations directes à partir du modèle discrétisé s'avérant, en général, biaisées lorsque le processus ne dispose pas d'une version discrète exacte, on est alors conduit à utiliser des méthodes indirectes comme la méthode par inférence indirecte (*cf.* GIET [2003]) ou encore la méthode des moments efficients pour estimer le modèle.

De plus, de nombreuses problématiques impliquent que l'on soit capable de simuler l'évolution de cours ou de taux modélisés par des processus continus ; la réalisation pratique de telles simulations nécessitera là encore la discrétisation de ces processus. C'est notamment le cas des problématiques de type DFA (*Dynamic Financial Analysis*, *cf.* KAUFMANN, GADMER et KLETT [2001] et HAMI [2003]) qui consistent en la modélisation de tous les facteurs ayant un impact sur les comptes d'une société d'assurance dans le but d'étudier sa solvabilité ou de déterminer une allocation d'actif optimale par exemple. Le nombre de variables à modéliser est alors très important : actifs financiers, sinistralité de chaque branche assurée, inflation, réassurance, concurrence et dépendances entre ces variables. Certaines de ces variables seront modélisées par des processus à diffusion ; lorsque l'équation différentielle stochastique (EDS) concernée dispose d'une solution explicite, comme c'est le cas pour le modèle de VASICEK [1977], la discrétisation s'impose à l'utilisateur. Lorsqu'une telle expression n'est pas disponible, l'utilisateur pourra se tourner notamment vers les schémas d'Euler ou de Milstein qui sont des développements d'Itô-Taylor à des ordres plus ou moins importants de l'EDS considérée. L'approximation sera d'autant plus satisfaisante, au sens du critère de convergence forte (*cf.* KLOEDEN et PLATEN [1995]), que l'ordre du développement est élevé.

Par ailleurs, la performance des simulations étant conditionnée par le générateur de nombres aléatoires utilisé, une attention particulière doit être portée à son choix. Nous comparons ici deux générateurs : le générateur pseudo-aléatoire *Rnd* d'Excel / Visual Basic et le générateur quasi-aléatoire obtenu à partir de l'algorithme de la translation irrationnelle du tore présenté par PLANCHET et JACQUEMIN [2003]. L'algorithme du tore s'avère nettement plus performant que *Rnd* mais souffre de la dépendance terme à terme des valeurs générées qui induit un biais dans la construction de trajectoires. Pour remédier à ce problème, nous proposons un générateur du *tore mélangé* qui conserve les bonnes propriétés de répartition de l'algorithme du tore et peut être utilisé pour construire des trajectoires de manière efficace.

L'objectif du présent article est de proposer au praticien quelques guides méthodologiques lui permettant d'effectuer de manière efficace des simulations de processus continus, plus

---

[1] Professeur associé de Finance et d'Assurance à l'ISFA et actuaire associé chez Winter & Associés, fplanchet@winter-associes.fr.

[2] Chargé de cours en assurance et étudiant en doctorat à l'ISFA et actuaire chez Winter & Associés, ptherond@winter-associes.fr.

Institut de Science Financière et d'Assurances (ISFA) – Université Lyon 1 - 50, avenue Tony Garnier - 69366 Lyon Cedex 07 - France.

Winter & Associés - 9, rue Beaujon - 75008 Paris et 18, avenue Félix Faure - 69007 Lyon - France.

particulièrement dans le contexte des problématiques d'assurance. Nous abordons donc les trois étapes clé que sont l'estimation des paramètres, la discrétisation du processus et la génération de nombres aléatoires.

## 2 Discrétisation de processus continus

Prenons le cas d'un processus défini par l'EDS suivante :

$$\begin{cases} dX_t = \mu(X_t, t)dt + \sigma(X_t, t)dB_t \\ X_0 = x \end{cases} \quad (1)$$

où $B$ est un mouvement brownien standard.

La mise en oeuvre pratique de ce processus va nécessiter la discrétisation de celui-ci. Pour cela on met l'équation (1) sous sa forme intégrale :

$$X_t = x + \int_0^t \mu(X_s, s)ds + \int_0^t \sigma(X_s, s)dB_s. \quad (2)$$

Si le processus considéré ne dispose pas d'une discrétisation exacte, un développement d'Itô-Taylor de l'équation (2) nous permet de disposer d'une version discrétisée approximative (voir le paragraphe 2.2). Cette approximation est d'autant plus précise que le développement intervient à un ordre élevé.

### 2.1 Discrétisation exacte

La simulation d'un processus d'Itô pourra être effectuée directement (sans erreur de discrétisation) dès lors que celui-ci admet une discrétisation exacte. Rappelons la définition d'une discrétisation exacte.

**Définition :** *Un processus* $\left(\widetilde{X}_{k\delta}\right)_{k \in [1; T/\delta]}$ *est une **discrétisation exacte** du processus $X$ si* $\forall \delta > 0, \forall k \in [1; T/\delta] \; \widetilde{X}_{k\delta} =_{loi} X_{k\delta}$.

Un processus $X$ admet une discrétisation exacte dès lors que l'on peut résoudre explicitement l'EDS qui lui est associée. C'est notamment le cas du mouvement brownien géométrique retenu par BLACK et SCHOLES [1973] pour modéliser le cours d'une action ou encore celui du processus d'Ornstein-Ulhenbeck retenu par VASICEK [1977] pour modéliser le taux d'intérêt instantané $r$ :

$$dr_t = a(b - r_t)dt + \sigma dB_t. \quad (3)$$

On rappelle que l'équation (3) exprimée dans l'univers historique ne suffit pas à définir le prix des zéro-coupons, et doit être complétée par la forme de la prime de risque. Celle-ci est choisie constante dans le modèle de VASICEK, ce qui permet de conserver la même forme de la dynamique dans l'univers corrigé du risque. L'équation (2) s'écrit dans ici :

$$r_t = r_0 e^{-at} + b(1 - e^{-at}) + \sigma e^{-at} \int_0^t e^{as} dB_s. \quad (4)$$

Les propriétés de l'intégrale d'une fonction déterministe par rapport à un mouvement brownien conduisent à la discrétisation exacte :

$$r_{t+\delta} = r_t e^{-a\delta} + b(1 - e^{-a\delta}) + \sigma \sqrt{\frac{1 - e^{-2a\delta}}{2a}} \varepsilon, \quad (5)$$

où $\varepsilon$ est une variable aléatoire de loi normale centrée réduite et $\delta$ est le pas de discrétisation retenu.

### 2.2 Discrétisation approximative

Lorsque la discrétisation exacte n'existe pas, il convient de se tourner vers des approximations discrètes du processus continu sous-jacent. Les schémas d'Euler et de Milstein sont les procédés de discrétisation les plus répandus. Tous deux sont des développements d'Itô-Taylor de l'équation (2) à des ordres différents (ordre 1 pour Euler, ordre 2 pour Milstein).

Dans la suite, nous ferons référence au critère de convergence forte pour classer les procédés de discrétisation. Une discrétisation approximative $\widetilde{X}$ converge fortement vers le processus continu $X$ lorsque l'erreur commise sur la valeur finale (à la date $T$) de la trajectoire obtenue par le processus discrétisé est en moyenne asymptotiquement négligeable, *i. e.* lorsque

$$\forall T > 0, \quad \lim_{\delta \to 0} \mathbf{E}\left[\left|\widetilde{X}_T^\delta - X_T\right|\right] = 0 \quad (6)$$

La vitesse de convergence de l'équation (6) nous permet d'introduire un ordre entre les procédés de discrétisation. Ainsi le processus discrétisé $\widetilde{X}$ converge fortement à l'ordre $\gamma$ vers le processus $X$ si :

$$\exists K > 0, \exists \delta_0 > 0, \forall \delta \in ]0, \delta_0] \quad \mathbf{E}\left[\left|\widetilde{X}_T^\delta - X_T\right|\right] \le K \delta^\gamma. \quad (7)$$

On notera que ce critère ne fait pas référence à la convergence uniforme.

*2.2.1 Schéma d'Euler*

Le procédé de discrétisation d'Euler consiste en l'approximation du processus continu $X$ par le processus discret $\widetilde{X}$ défini, avec les mêmes notations que précédemment, par :

$$\widetilde{X}_{t+\delta} = \widetilde{X}_t + \mu(\widetilde{X}_t, t)\delta + \sigma(\widetilde{X}_t, t)\sqrt{\delta}\varepsilon. \quad (8)$$

KLOEDEN et PLATEN [1995] prouvent que sous certaines conditions de régularité, le schéma d'Euler présente un ordre de convergence forte de 0,5.

Par exemple dans le modèle de Cox, Ingersoll et Ross (CIR), le taux d'intérêt instantané est solution de l'EDS :

$$dr_t = a(b - r_t)dt + \sigma\sqrt{r_t}\,dB_t. \quad (9)$$

Aussi le processus discret $\widetilde{r}$ déterminé par le schéma d'Euler peut s'écrire :

$$\widetilde{r}_{t+\delta} = \widetilde{r}_t + a(b - \widetilde{r}_t)\delta + \sigma\sqrt{\widetilde{r}_t * \delta}\,\varepsilon. \quad (10)$$

*2.2.2 Schéma de Milstein*

Le schéma de Milstein est obtenu en allant plus avant dans le développement d'Itô-Taylor. Le processus discret $\widetilde{X}$ est alors défini par :

$$\begin{aligned} \widetilde{X}_{t+\delta} = \widetilde{X}_t &+ \mu(\widetilde{X}_t, t)\delta + \sigma(\widetilde{X}_t, t)\sqrt{\delta}\varepsilon \\ &+ \frac{\sigma_x(\widetilde{X}_t, t)\sigma(\widetilde{X}_t, t)}{2}\delta(\varepsilon^2 - 1) \end{aligned} \quad (11)$$

où $\sigma_x(\widetilde{X}_t, t)$ désigne la dérivée par rapport au premier argument de la fonction $\sigma(.,.)$ évaluée en $(\widetilde{X}_t, t)$.

Ce procédé de discrétisation présente, en général, un ordre de convergence forte de 1. Remarquons que si la volatilité $\sigma$ est une constante, les procédés de discrétisation d'Euler et de Milstein conduiront à la même discrétisation, c'est le cas pour les modèles de Vasicek et de Hull et White (*cf.* HULL [1999]). En revanche pour le modèle de CIR, il vient :

$$\widetilde{r}_{t+\delta} = \widetilde{r}_t + a(b - \widetilde{r}_t)\delta + \sigma\sqrt{\widetilde{r}_t * \delta}\,\varepsilon + \frac{\sigma^2}{4}\delta(\varepsilon^2 - 1). \qquad (12)$$

En développant à des degrés supérieurs l'équation (2), il est possible d'obtenir des processus discrétisés d'ordre de convergence plus élevé. Toutefois ils nécessiteront des calculs plus nombreux et peuvent faire intervenir plus d'une variable aléatoire ce qui signifie des temps de simulation plus importants.

En outre, le schéma de Milstein pose des problèmes pratiques de mise en œuvre dès lors que l'on s'intéresse à des phénomènes de dimension supérieure ou égale à 2, *i. e.* dès que l'on cherche à simuler des vecteurs aléatoires.

Le graphique suivant permet de comparer les évolutions moyennes de la diffusion définie par (3) selon le schéma de discrétisation retenu pour les paramètres :

$r_0 = 4\%$ $\qquad b = 5\%$ $\qquad a = 0{,}5$ $\qquad \sigma = 10\%$

Ce graphique a été obtenu en faisant la moyenne de 10 000 simulations du taux d'intérêt générées.

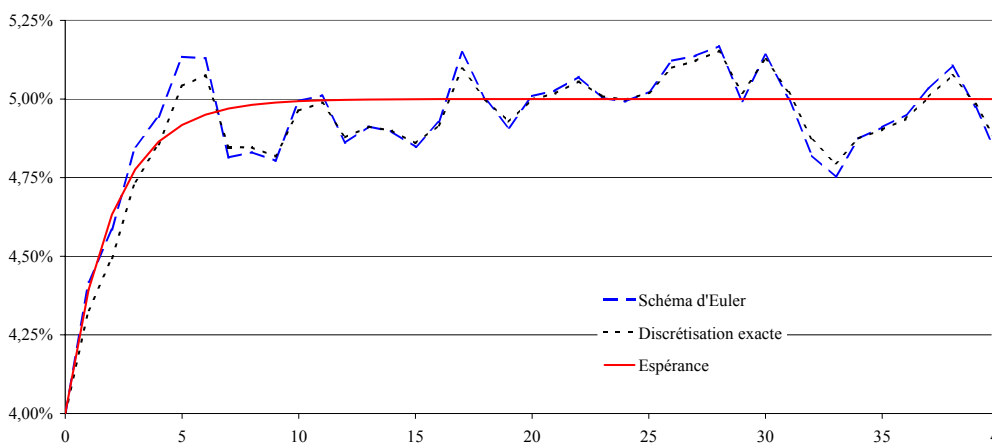

**Fig. 1.** Evolution moyenne du taux modélisé par Vasicek selon le procédé de discrétisation retenu

Les discrétisations exacte et selon le schéma d'Euler sont relativement proches graphiquement. L'écart est en particulier négligeable lorsqu'on s'intéresse aux évolutions moyennes d'un bon de capitalisation de prix $BC_t$ à l'instant $t$, qui évolue selon la dynamique $BC_{t+\delta} = BC_t\, e^{\delta r_t}$. A partir des taux simulés pour effectuer le graphique 1, on obtient le graphique suivant.

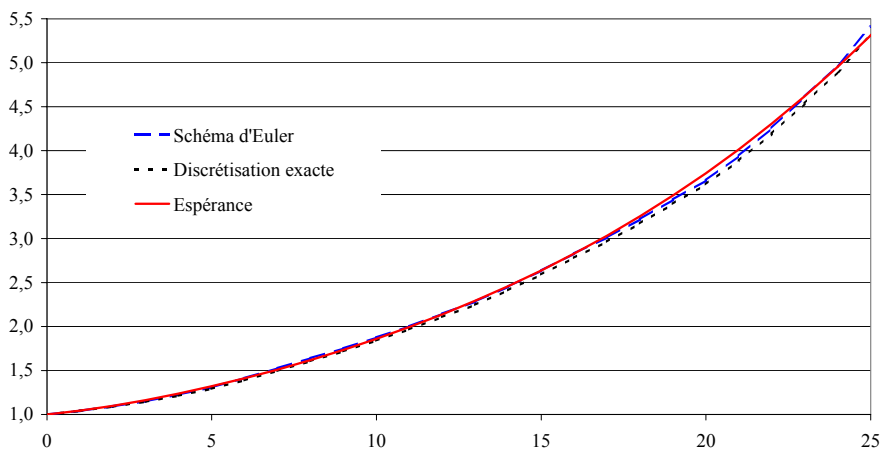

**Fig. 2.** Evolution moyenne d'un bon de capitalisation dont le taux évolue selon le modèle de Vasicek selon le procédé de discrétisation retenu

# 3   Estimation des paramètres

L'estimation des paramètres est une étape délicate dans la simulation de trajectoires d'un processus continu car elle peut être l'origine d'un biais. En effet, le praticien peut se voir confronté à deux problèmes :
- le processus n'admet pas forcément de discrétisation exacte,
- la variable modélisée n'est pas toujours directement observable.

En effet si le processus considéré n'admet pas de discrétisation exacte, il sera impossible d'estimer les paramètres du modèle par la méthode du maximum de vraisemblance. Par ailleurs il est souvent impossible d'exprimer de manière exacte les densités transitoires entre deux observations. Il faudra alors se tourner vers des méthodes simulées telles que l'inférence indirecte pour estimer les paramètres.

## 3.1   Le biais associé à la procédure d'estimation

On reprend ici la cas simple de la modélisation du taux à court terme par un processus d'Ornstein-Ulhenbeck, comme dans l'équation (3) ; la discrétisation exacte (5) conduit à un processus auto-régressif d'ordre 1, AR(1).

L'estimation des paramètres du modèle s'effectue classiquement en régressant la série des taux courts sur la série décalée d'une période, que l'on écrit sous la forme usuelle :

$$Y = \alpha + \beta X + \sigma_1 \varepsilon, \qquad (13)$$

avec

$$a = -\ln\beta, \; b = \frac{\alpha}{1-\beta} \; \text{et} \; \sigma^2 = \sigma_1^2 \frac{2\ln\beta}{\beta^2 - 1}. \qquad (14)$$

L'estimateur du maximum de vraisemblance (EMV) coïncide avec l'estimateur des moindres carrés ; on obtient ainsi les estimateurs suivants des paramètres du modèle d'origine :

$$\hat{a}_{exact} = -\ln\hat{\beta}, \; \hat{b}_{exact} = \frac{\hat{\alpha}}{1-\hat{\beta}} \; \text{et} \; \hat{\sigma}^2_{exact} = \hat{\sigma}^2 \frac{2\ln\hat{\beta}}{\hat{\beta}^2 - 1}. \qquad (15)$$

On constate en particulier que ces estimateurs, s'ils sont EMV, sont biaisés. Ceci peut être gênant, notamment dans le cas où les paramètres du modèle ont une interprétation dans le modèle, comme par exemple $b$ dans le cas présent, valeur limite du taux court. On voit sur cet exemple qu'on sera donc conduit à mettre en œuvre des procédés de réduction du biais.

## 3.2   Principe de l'estimation par inférence indirecte

Cette méthode est utilisée lorsque le processus n'admet pas de discrétisation exacte ou que sa vraisemblance, trop complexe, ne permet pas d'implémenter la méthode du maximum de vraisemblance. Elle consiste à choisir le paramètre $\hat{\theta}$ qui minimise (dans une métrique à définir) la distance entre l'estimation d'un modèle auxiliaire sur les observations et l'estimation de ce même modèle auxiliaire sur les données simulées à partir du modèle de base pour $\theta = \hat{\theta}$. Le modèle auxiliaire étant une discrétisation approximative de (1), le schéma d'Euler est souvent utilisé pour servir de modèle auxiliaire.

Cette méthode conduit à des estimations bien plus précises que celles obtenues à partir d'estimations « naïves » des paramètres du modèle auxiliaire sur les données observées.

GIET [2003] étudie l'impact du choix du procédé de discrétisation utilisé pour fournir le modèle auxiliaire sur l'estimation par inférence indirecte et montre qu'utiliser un procédé d'ordre de convergence plus élevé (comme la schéma de Milstein par rapport au schéma d'Euler) permet de réduire considérablement le biais sur l'estimation des paramètres.

Toutefois, que ce soit dans le cas d'une estimation par maximum de vraisemblance ou par inférence indirecte, on doit observer que dans le cas d'un modèle de taux tel que celui de Vasicek, l'estimation des paramètres de la dynamique d'évolution du taux court dans l'univers historique ne suffit pas et doit être complétée par l'estimation de la prime de risque, qui constitue un point délicat (*cf.* LAMBERTON et LAPEYRE [1997] sur ce sujet).

## 3.3   Les méthodes d'estimation *ad hoc*

En pratique, dans les problèmes assurantiels, la grandeur modélisée par un processus continu n'est, le plus souvent, pas la grandeur d'intérêt : typiquement, on modélise le taux court, par exemple par un processus de diffusion, mais dans le cadre d'une problématique d'allocation d'actif pour un régime de rentiers, il sera déterminant que le modèle représente correctement les prix des obligations d'échéances longues. De manière équivalente, on est amené à estimer les paramètres du modèle pour représenter correctement la courbes des taux zéro-coupon, qui est dans une relation bijective avec la courbe du prix des obligations zéro-coupon :

$$P(0,T) = \exp\{-T R(0,T)\}. \qquad (16)$$

L'idée est alors d'estimer les paramètres pour minimiser une distance (par exemple la distance quadratique) entre les prix prédits par le modèle et les prix observés sur le marché. Cette méthode est notamment utilisée dans FARGEON et NISSAN [2003].

De plus, en pratique on pourra s'interroger sur les paramètres que l'on estime et les paramètres que l'on fixe arbitrairement compte tenu de la connaissance que l'on peut avoir du contexte par ailleurs. Ainsi, dans l'exemple évoqué, l'estimation simultanée des paramètres $a$, $b$ et $\sigma$ conduit à une courbe des taux quasi déterministe ($\sigma$ est petit), ce qui peut apparaître irréaliste. On est ainsi conduit à fixer arbitrairement $\sigma$, puis à estimer les deux paramètres restants.

Dans cette approche il convient également d'être attentif au choix effectué quand à la paramétrisation du modèle : courbe des taux zéro-coupon ou courbe des prix des zéro-coupon ; en effet, la fonction de correspondance entre ces deux grandeurs accentue les écarts de courbure et le choix de l'une ou l'autre des paramétrisations peut conduire à des résultats sensiblement différents. Ce point est développé sur le plan théorique par RONCALLI [1998] et est illustré sur un exemple dans la section suivante.

Au surplus, on notera que l'exploitation directe de données telles que le prix des zéro-coupon (ou les taux zéro-coupon) évite l'estimation de la prime de risque ; celle-ci est en effet incluse dans ces prix de marché et conduit ainsi à travailler naturellement dans l'univers corrigé du risque. L'estimation directe des paramètres sur des historiques de taux court nécessite, on l'a vu, l'estimation (délicate) de la prime de risque. Ce point est illustré par exemple dans LAMBERTON et LAPEYRE [1997].

Dans les illustration présentées ci-après, l'exploitation de données incluant les primes de risque est privilégiée.

### 3.4 Illustration dans le cas du modèle de Vasicek

A partir de la courbe des taux publiée par l'Institut des Actuaires, nous avons estimé les paramètres du modèle de Vasicek selon l'approche *ad hoc* décrite *supra*. L'estimation est menée d'une part directement à partir des valeurs des taux zéro-coupon de la courbe et d'autre part à partir des prix des obligations qui s'en déduisent. Cette courbe des taux est construite à partir des prix de marché et elle inclut les primes de risque.

Estimation *ad hoc* sur le prix des zéro-coupons : cette procédure consiste à déterminer par une méthode de type « moindres carrés » les paramètres du modèle de Vasicek qui permettent de représenter au mieux les prix des zéro-coupons. Cette méthode a été mise en œuvre en estimant les trois paramètres a, $\tilde{b}$ et σ puis en fixant σ et en n'estimant plus que le deux autres paramètres.

Comme on peut le voir dans le tableau 1, l'estimation des paramètres du modèle de Vasicek donne des résultats très différents selon l'approche retenue :

| Paramètres estimés | EMC sur le prix des zéro-coupon | | | EMC sur les taux zéro-coupon | | |
|---|---|---|---|---|---|---|
| | a, b et σ | a et b | a et b | a, b et σ | a et b | a et b |
| **a =** | *0,2293* | *0,2353* | *0,5032* | *0,3951* | *0,3253* | *0,4232* |
| **b =** | *0,0572* | *0,0760* | *0,0701* | *0,0624* | *0,0745* | *0,0869* |
| **r0 =** | 0,0207 | 0,0207 | 0,0207 | 0,0207 | 0,0207 | 0,0207 |
| **σ =** | *0,0000* | 0,0500 | 0,1000 | *0,0000* | 0,0500 | 0,1000 |

**Tab. 1.** Paramètres du modèle de Vasicek estimés selon les techniques d'estimation des paramètres

Le prix $P(t)$ en 0 d'un zéro-coupon d'échéance $t$ est donné par la formule classique :

$$P(t) = \exp\left\{-t*\left[\tilde{b} - \frac{\sigma^2}{2a} - \frac{1}{at}\left(\left(\tilde{b} - \frac{\sigma^2}{2a} - r_0\right)\left(1 - e^{-at}\right) - \frac{\sigma^2}{4a}\left(1 - e^{-at}\right)^2\right)\right]\right\} \quad (17)$$

La figure 3, qui indique le prix des zéro-coupons, en fonction de leur échéance, selon les différentes méthodes d'estimation des paramètres, nous permet d'observer que l'estimation des paramètres par le critère des moindres carrés sur les taux zéro-coupon s'avère moins performant que la méthode *ad hoc* si la variable d'intérêt est le prix des obligations zéro-coupon.

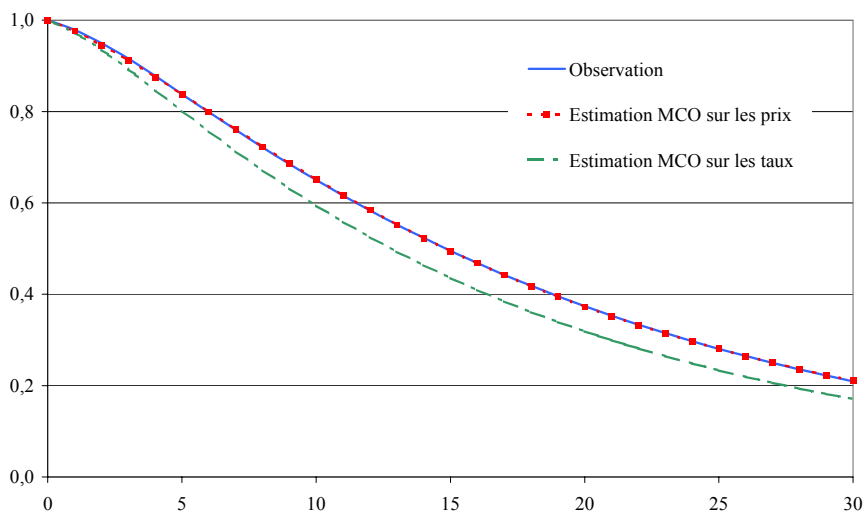

**Fig. 3.** Prix des zéro-coupons en fonction de leur échéance et de la méthode d'estimation des paramètres

En revanche, cette technique d'estimation des paramètres permet de générer des taux zéro-coupon nettement plus proches en espérance de la courbe des taux originelle que ceux obtenus par la méthode *ad hoc* :

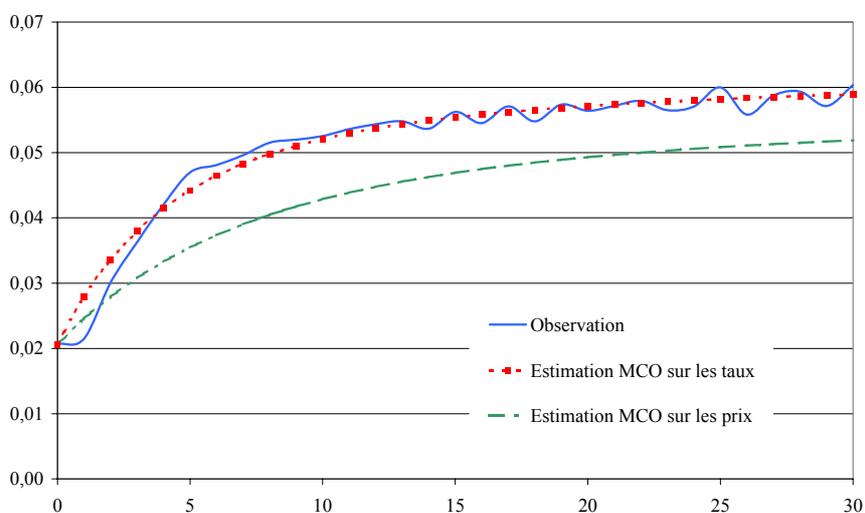

**Fig. 4.** Taux instantanés espérés en fonction de leur échéance et de la technique d'estimation des paramètres

Si les taux obtenus par la méthode *ad hoc* sur le prix des zéro-coupons sont, en espérance, éloignés des taux observés, ils permettent néanmoins d'approcher le prix des zéro-coupons avec une bonne précision.

Ainsi, en pratique, on sera conduit à privilégier pour une problématique d'allocation d'actifs d'un régime de rentes l'approche *ad hoc* sur le prix des zéro-coupons.

## 4  Génération des trajectoires

La génération des trajectoires passe nécessairement par la génération de nombres aléatoires. De manière pratique, il s'agit de générer des réalisations de variable aléatoire de loi uniforme sur le segment [0, 1]. En effet, si $u$ est une telle réalisation, $F^{-1}(u)$ peut s'apparenter à une réalisation d'une variable aléatoire de fonction de répartition $F$. La technique d'inversion de la fonction de répartition permet ainsi à partir de réalisations de variables uniformes, d'obtenir de simuler des réalisations d'autres variables aléatoires. Lorsqu'on ne dispose pas de formule explicite pour $F^{-1}$, on utilisera des algorithmes d'approximation de cette fonction ou des algorithmes spécifiques à la loi que l'on souhaite traiter Par exemple, pour simuler une réalisation d'une loi de Poisson de paramètre $\lambda$, on générera une suite $(V_i)$ de réalisations de variable aléatoire exponentielle de paramètre 1. En effet $N = \mathbf{Inf}\left\{ n \left| \sum_{i=1}^{n+1} V_i > \lambda \right. \right\}$ suit une loi de Poisson de paramètre $\lambda$.

Les modélisations retenues en finance et en assurance faisant souvent intervenir des mouvements browniens, il est nécessaire de simuler des réalisations de variables aléatoires N(0,1). On ne dispose pas de formule exacte de l'inverse de la fonction de répartition inverse de la loi normale centrée réduite, mais l'algorithme de Box-Muller permet à partir de deux variables uniformes indépendantes sur [0,1] de générer deux variables indépendantes N(0,1). En effet, si $U_1$ et $U_2$ sont des v.a. U[0,1] indépendantes alors en posant :

$$\begin{cases} X_1 = \sqrt{-2\ln U_1}\cos(2\pi U_2) \\ X_2 = \sqrt{-2\ln U_1}\sin(2\pi U_2) \end{cases} \quad (18)$$

Les variables aléatoires $X_1$ et $X_2$ sont indépendantes et suivent une loi gaussienne centrée réduite. Cette technique (exacte) est toutefois relativement longue à mettre en œuvre et nécessite l'indépendance des réalisations uniformes générées. AUGROS et MORÉNO [2002] présentent divers algorithmes pour approximer l'inverse de la fonction de répartition de la loi normale centrée réduite. Dans la suite, nous utiliserons l'algorithme de Moro (*cf.* PLANCHET et JACQUEMIN [2003]) qui allie rapidité et précision.

### 4.1 Deux générateurs de nombres aléatoires : *Rnd* et l'algorithme du tore

*4.1.1 Le générateur implémenté dans Excel / Visual Basic*

Le générateur implémenté dans Excel (*Rnd*) est un générateur congruentiel, c'est à dire un générateur périodique issu d'une valeur initiale (on parle également de « graine » du générateur). Changer de valeur initiale permet de changer de suite de nombres. Le lecteur se référera à PLANCHET et JACQUEMIN [2003] pour plus d'informations sur l'implémentation informatique de tels générateurs.

*4.1.2 La translation irrationnelle du tore*

Ce générateur multidimensionnel donne à la $n$-ème réalisation de la $d$-ème variable aléatoire uniforme à simuler la valeur $u_n$ :

$$u_n = n\sqrt{p_d} - \left[ n\sqrt{p_d} \right], \quad (19)$$

où $p_d$ est le $d$-ème nombre premier et [.] désigne l'opérateur partie entière.

### 4.1.3 Quelques éléments de comparaison

Les tests statistiques élémentaires (test d'adéquation du $\chi^2$, de Kolmogorov-Smirnov et d'Anderson-Darling) rappelés en annexe permettent d'apprécier la supériorité de répartition de la suite générée par le tore par rapport à celle générée par *Rnd*.

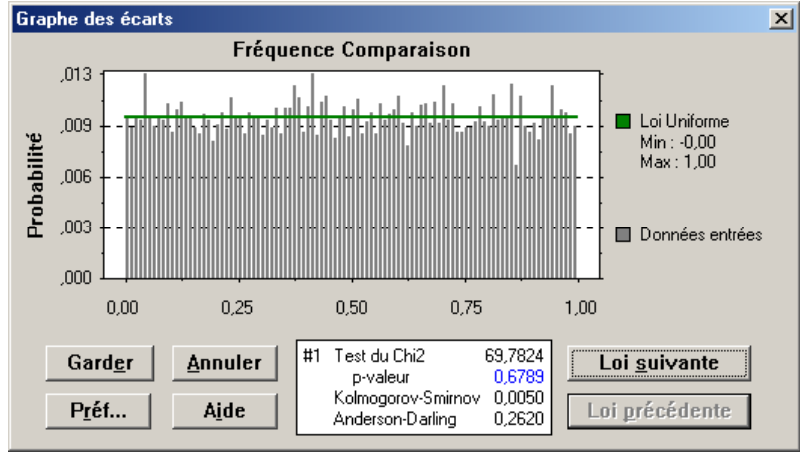

**Fig. 5.** Répartition et tests statistiques du générateur Rnd

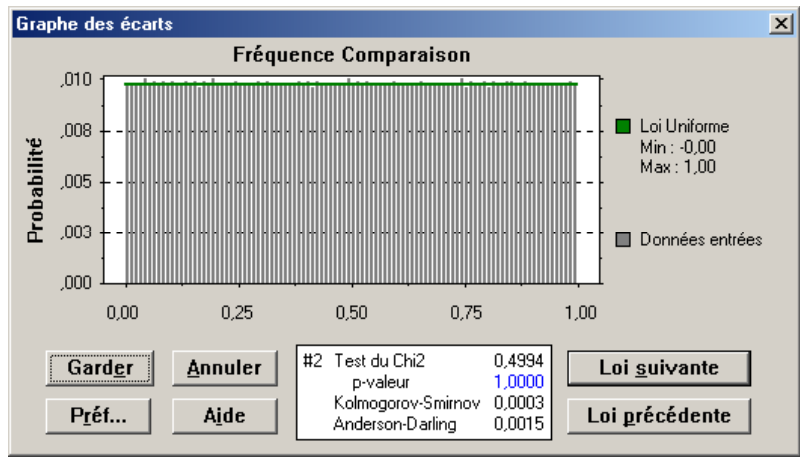

**Fig. 6.** Répartition et tests statistiques du générateur du tore*

Par ailleurs, pour comparer les performances des deux générateurs, nous avons étudié la précision de l'estimation empirique du prix d'un call européen sur une action modélisée par un mouvement brownien géométrique sous la probabilité risque-neutre :

$$dS_t = rS_t dt + \sigma S_t d\widetilde{B}_t, \qquad (20)$$

où $r$ est le taux sans risque et $\widetilde{B}$ un mouvement brownien sous la mesure risque-neutre. Rappelons que, sous les hypothèses du modèle de Black et Scholes, la probabilité risque-neutre est l'unique probabilité sous lesquels les prix actualisés sont des martingales. La densité de Radon-Nikodym de cette probabilité par rapport à la probabilité historique est donnée par le théorème de Girsanov (*cf.* AUGROS et MORÉNO [2002]).

Ce processus dispose d'une discrétisation exacte :

$$S_{t+\delta} = S_t \exp\left\{\left(r - \frac{\sigma^2}{2}\right)\delta + \sigma\sqrt{\delta}\,\varepsilon\right\}, \qquad (21)$$

où $\varepsilon$ est une variable aléatoire de loi N(0,1).

Par ailleurs, nous disposons d'une formule fermée pour le prix en $t$ d'un Call européen de prix d'exercice $K$ et d'échéance $T$ sur ce titre :

$$C_t(S, K, T) = S_t N(d_1) - K e^{-r(T-t)} N(d_2), \qquad (22)$$

où :

- $d_1 = \dfrac{\ln \dfrac{S_t}{K} + \left(r + \dfrac{\sigma^2}{2}\right)(T-t)}{\sigma\sqrt{T-t}}$,

- $d_2 = d_1 - \sigma\sqrt{T-t}$,
- $N$ désigne la fonction de répartition de la loi normale centrée réduite.

Nous pouvons donc mesurer la performance des générateurs par le biais de l'erreur d'estimation du prix de l'option considérée. En effet, si $S_T^i$ est la prix du titre considéré à la date $T$ dans la $i$-ème simulation, l'estimateur naturel du prix de l'option en $t$ est :

$$\hat{C}_t(S,K,T,n) = \frac{e^{-r(T-t)}}{n}\sum_{i=1}^{n}\left[s_T^i - K\right]^+. \qquad (23)$$

L'erreur relative d'estimation peut donc s'écrire :

$$\rho = \frac{\hat{C}_t(S,K,T,n) - C_t(S,K,T)}{C_t(S,K,T)}. \qquad (24)$$

Le graphique suivant présente l'évolution de $\rho$ selon le nombre de simulations effectuées pour une action de volatilité 20 % et un Call d'échéance 6 mois.

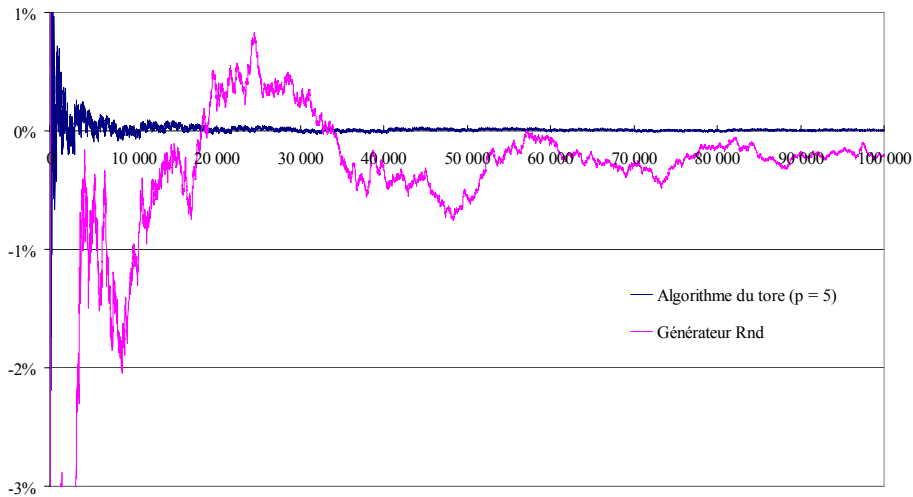

**Fig. 7.** Erreur d'estimation en fonction du nombre de simulations

L'estimation à partir des valeurs générées par *Rnd* est biaisée de manière systématique de prés de 0,2 % alors celle effectuée à l'aide du tore converge rapidement vers la valeur théorique : à partir de 7 000 simulations, l'erreur relative est inférieure à 0,1 %.

Il ressort des ces différentes comparaisons que le générateur du tore semble plus performant que *Rnd*. Son utilisation connaît toutefois un certain nombre de limites.

### 4.2 Limites de l'algorithme du tore

Les valeurs générées par le tore ne sont pas indépendantes terme à terme, ceci peut générer des erreurs non négligeables. Par exemple nous avons généré 10 000 trajectoires, avec un pas de discrétisation de 1, sur 20 ans d'un mouvement brownien géométrique de volatilité 20 % et avons comparé l'évolution moyenne du cours estimée à partir des simulations et l'évolution moyenne théorique. En effet, si $s_t^i$ est le cours du titre à la date t dans la $i$-ème simulation, l'estimateur empirique du cours moyen à la date $t$, $\bar{s}_t$ est donné par :

$$\bar{s}_t = \frac{1}{n}\sum_{i=1}^{n} s_t^i. \qquad (25)$$

Le graphique suivant présente les deux trajectoires obtenues.

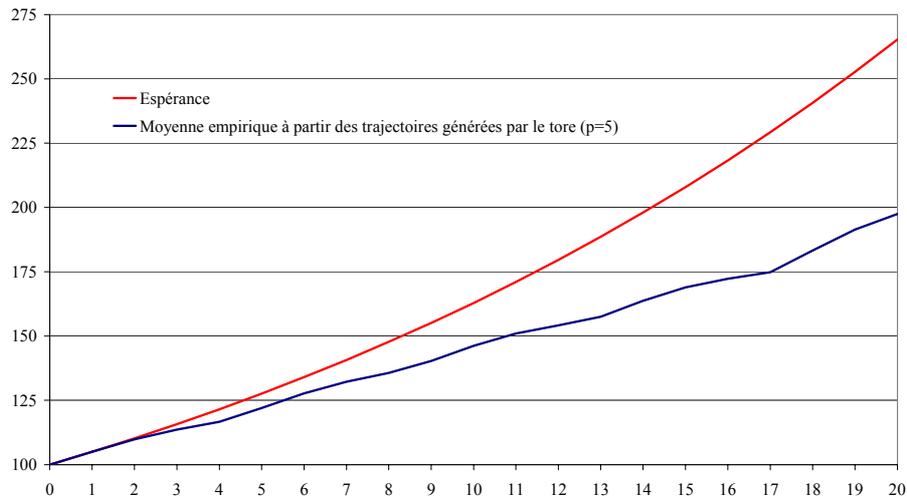

**Fig. 8.** Trajectoires espérée et moyenne des 10 000 trajectoires simulées par le tore d'un mouvement brownien géométrique

A partir de *t* = 2, les trajectoires générées à partir de l'algorithme du tore sont en moyenne très en dessous de la trajectoire moyenne. Cet algorithme n'est donc pas utilisable en l'état pour générer des trajectoires. Ce biais s'explique par la dépendance terme à terme des valeurs générées par le tore que nous permet d'observer le corrélogramme de la suite générée par ce générateur.

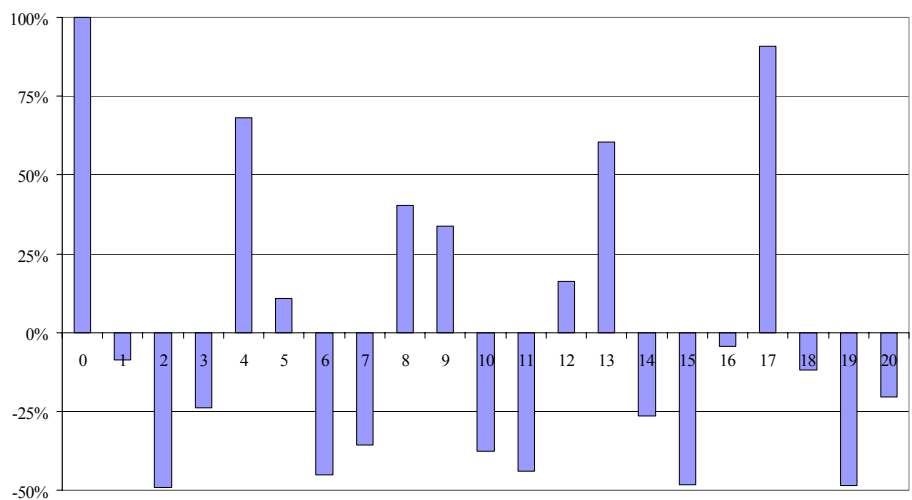

**Fig. 9.** Corrélogramme de la suite générée par le tore (*p* = 5)

Rappelons que le *h*-ème terme du corrélogramme $\rho_h$ s'écrit :

$$\rho_h = \frac{\sum_{k=1}^{n}(u_k - \bar{u})(u_{k+h} - \bar{u})}{\sum_{k=1}^{n}(u_k - \bar{u})^2}, \qquad (26)$$

où $\bar{u}$ désigne la moyenne empirique de la suite *u*.

Le « test du poker » permet également de mettre en évidence cette faille de l'algorithme du tore. L'idée de ce test est de comparer les fréquences théoriques des mains au poker avec les fréquences observées sur les simulations effectuées. De manière pratique, ce test consiste à prendre des listes de quatre chiffres tirés aléatoirement, de manière à observer les « combinaisons de valeurs » puis à effectuer un test d'adéquation du $\chi^2$ pour voir si les fréquences observées correspondent aux fréquences théoriques. On distingue ainsi cinq cas :
- les quatre chiffres sont tous différents,
- la liste contient une et une seule paire,
- la liste contient deux paires,
- la liste contient un brelan (trois chiffres identiques),
- la liste contient un carré (quatre chiffres identiques).

En constituant les listes en prenant les chiffres dans l'ordre où ils sont simulés, le tableau 2 résume les fréquences suivantes.

| | Fréquences théorique | Fréquences observées pour l'algorithme du tore | | | | | | | | |
|---|---|---|---|---|---|---|---|---|---|---|
| | | p = 2 | p = 3 | p = 5 | p = 7 | p = 11 | p = 13 | p = 17 | p = 19 | p = 23 | p = 29 |
| Carré | 0,1% | 0% | 0% | 0% | 0% | 0% | 0% | 0% | 0% | 0% |
| Brelan | 3,6% | 0% | 0% | 0% | 0% | 0% | 0% | 0% | 0% | 0% |
| Double paire | 2,7% | 0% | 0% | 0% | 0% | 0% | 0% | 0% | 0% | 0% |
| Paire | 43,2% | 0% | 0% | 0% | 37,2% | 49,7% | 0% | 0% | 23,3% | 0% | 0% |
| Tous différents | 50,4% | 100% | 100% | 100% | 62,8% | 50,3% | 100% | 100% | 76,7% | 100% | 100% |
| p-valeur du χ2 | 1,00 | 0,91 | 0,91 | 0,91 | 1,00 | 1,00 | 0,91 | 0,91 | 0,99 | 0,91 | 0,91 |

**Tab. 2.** Test du poker sur 4 000 réalisations générées par l'algorithme du tore

Les p-valeurs des tests d'adéquation du $\chi^2$ sont proches de 1, toutefois quel que soit $p$, le tore ne conduit jamais à l'obtention d'un carré, ni d'un brelan, ni d'une double paire alors que ces trois situations représentent 6,4 % des cas.

### 4.3 Générateur du *tore mélangé*

Pour contourner le problème posé par la dépendance terme à terme des valeurs générées par l'algorithme du tore, nous proposons une adaptation qui consiste à « mélanger » ces valeurs avant de les utiliser.

#### 4.3.1 Descriptif de l'algorithme

Notons $(u_n)$ la suite générée par le nombre premier $p$. Au lieu d'utiliser la nombre $u_n$ lors du $n$-ème tirage de la loi uniforme sur [0, 1], nous proposons d'utiliser $u_m$ où $m$ est choisi de manière aléatoire dans **N**.

Le générateur ainsi obtenu présente les mêmes bonnes caractéristiques globales que l'algorithme du tore sans la dépendance terme à terme. Il nécessite toutefois davantage de temps de simulation du fait du tirage de l'indice $m$. Nous proposons l'algorithme suivant lorsque l'on souhaite générer $N$ réalisation de variables uniformes :

$$u_m = u_{\varphi(n)}, \quad (27)$$

où :

$$\varphi(n) = [\alpha * N * \tilde{u} + 1], \quad (28)$$

où :
- [.] désigne l'opérateur partie entière,
- $\alpha \geq 1$,
- $\tilde{u}$ est la réalisation d'une variable aléatoire de loi uniforme.

Le facteur $\alpha$ de l'équation (28) a pour vocation de réduire le nombre de tirages qui donneraient lieu au même indice et donc au même nombre aléatoire. En effet, plus $\alpha$ est grand plus la probabilité de tirer deux fois le même nombre aléatoire est faible. Dans la pratique, $\alpha = 10$ est satisfaisant.

#### 4.3.2 Choix de la procédure de mélange

Pour la génération de $\tilde{u}$, nous avons retenu le générateur *Rnd* ou tout autre générateur congruentiel ayant une période importante. En effet, utiliser l'algorithme du tore pour effectuer le mélange ne réduit pas la corrélation observée comme le montre le corrélogramme suivant.

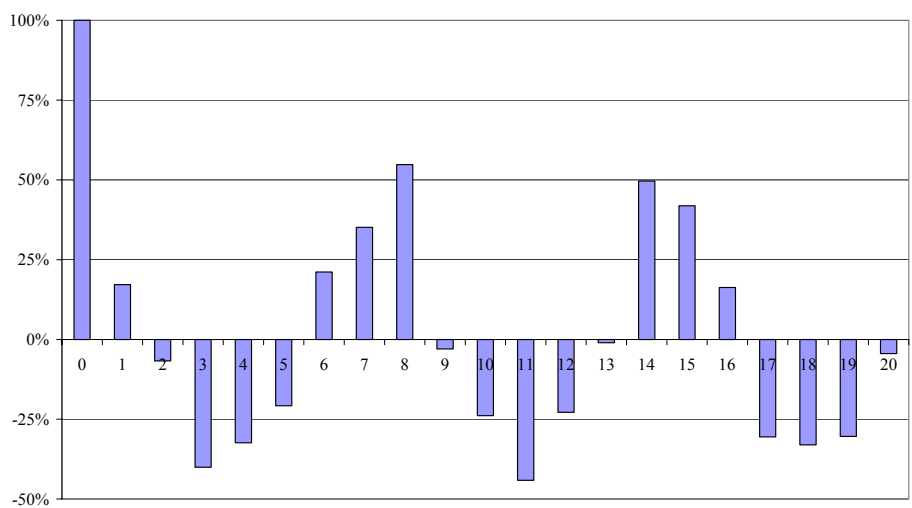

**Fig. 10.** Corrélogramme de la suite générée par le tore ($p_1$ = 5) mélangé par le tore ($p_2$ = 19)

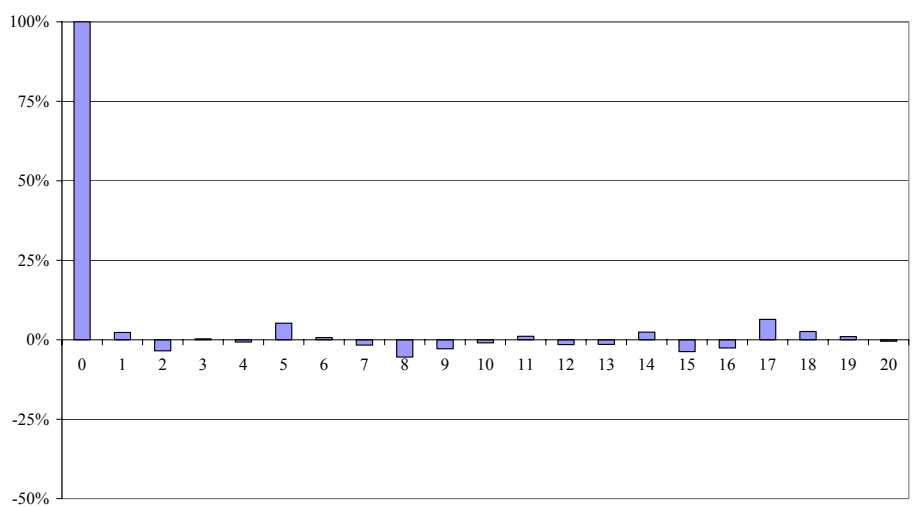

**Fig. 11.** Corrélogramme de la suite générée par le tore ($p_1 = 5$) mélangé par *Rnd*

En revanche l'utilisation de *Rnd* permet de réduire considérablement les corrélations. La figure 11 nous permet de constater que le mélange a pratiquement fait disparaître la corrélation terme à terme, le générateur du *tore mélangé* conservant néanmoins les propriétés de bonne répartition globale et de rapidité de convergence du tore.

De plus, comme le montre le tableau 3, le tore mélangé par *Rnd* satisfait de manière satisfaisante le test du poker puisque toutes les « mains » sont représentées dans des proportions proches des fréquences théoriques.

|  | Fréquences théorique | Fréquences observées pour le tore mélangé par Rnd (p=2) |
|---|---|---|
| **Carré** | 0,1% | 0,1% |
| **Brelan** | 3,6% | 2,8% |
| **Double paire** | 2,7% | 2,5% |
| **Paire** | 43,2% | 37,2% |
| **Tous différents** | 50,4% | 57,4% |
| **p-valeur du $\chi 2$** | 1,00 | 1,00 |

**Tab. 3.** Test du poker sur 40 000 réalisations générées par le tore mélangé par Rnd

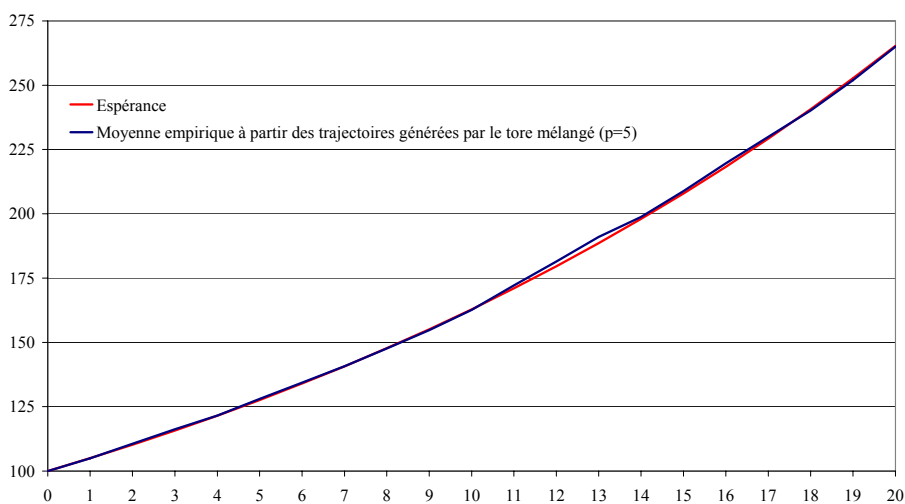

**Fig. 12.** Trajectoires espérée et moyenne des 5 000 trajectoires simulées
par le tore mélangé d'un mouvement brownien géométrique

La figure 12 nous permet de vérifier que les trajectoires générées à partir du tore mélangé sont satisfaisantes.

## 5 Conclusion

Dans cet article nous avons présenté les trois étapes indispensables à la génération pratique de trajectoires de variables modélisées par des processus continus : l'estimation des paramètres, la discrétisation du processus et la génération de nombres aléatoires.

L'étape de discrétisation conduit à arbitrer entre précision et temps de calcul à moins qu'une discrétisation exacte peu coûteuse en calculs soit disponible. Lorsque ce n'est pas le cas, on préférera le schéma de Milstein au schéma d'Euler car il demande le même nombre de tirages de nombres aléatoires pour une meilleure précision.

L'estimation des paramètres doit faire l'objet d'une attention particulière du fait des biais auxquels conduirait une estimation « naïve ». Il convient de préférer la méthode d'estimation par inférence indirecte à une estimation directe souvent porteuse de biais. Le principe de l'estimation *ad hoc* permettra également d'estimer des paramètres satisfaisants lorsque la variable d'intérêt n'est pas directement la variable simulée, ce qui est le plus souvent le cas dans les problématiques d'assurance.

Enfin nous recommandons l'utilisation du *tore mélangé* pour toute construction pas-à-pas de trajectoires. Ce générateur, simple à mettre en place, donne des résultats très satisfaisants, notamment comparé au *Rnd* d'Excel.

## Annexe : Tests d'adéquation à une loi

Le lecteur trouvera un descriptif plus détaillé de ces tests dans SAPORTA [1990] pour les deux premiers et dans PARTRAT et BESSON [2004] pour le dernier.

### Test du $\chi 2$

Soit $X$ une variable aléatoire à valeurs dans **E** de loi $\mathbf{P}_X$ inconnue et $\mathbf{P_0}$ une loi connue sur **E**. Soit $A_1, \ldots, A_d$ une partition de **E** telle que $\pi_k = \mathbf{P_0}[A_k] > 0$ pour tout $k$. Nous disposons d'un $n$-échantillon indépendant et identiquement distribué $(X_1, \ldots, X_n)$ de $X$. Soit $N_k$ le nombre de variables aléatoires $X_i$ dans $A_k$. Si $\mathbf{P}_X = \mathbf{P}_0$ ($H_0$) considérons la statistique $D^2$ définie comme suit :

$$D^2 = \sum_{k=1}^{d} \frac{(N_k - n\pi_k)^2}{n\pi_k}. \qquad (29)$$

$D^2$ est asymptotiquement distribuée comme une variable de $\chi^2_{d-1}$. La p-valeur de ce test est donc donnée par :

$$\hat{\alpha} = \mathbf{P}\left[\chi^2_{d-1} > D^2\right]. \qquad (30)$$

### Test de Kolmogorov-Smirnov

Si $F_n^*$ est la fonction de répartition empirique d'un $n$-échantillon d'une variable aléatoire de fonction de répartition $F$, alors la statistique $D_n = \mathbf{Sup}\left|F_n^*(x) - F(x)\right|$ est asymptotiquement distribuée comme suit :

$$\mathbf{P}\left[\sqrt{n}\, D_n < y\right] \to \sum_{-\infty}^{\infty} (-1)^k \exp\left\{-2k^2 y^2\right\}. \qquad (31)$$

La statistique $D_n$ est indépendante de $F$ et le résultat (31) nous permet de tester :

$$\begin{cases} H_0 : F(x) = F_0(x) \\ H_1 : F(x) \neq F_0(x) \\ \alpha \end{cases} \qquad (32)$$

### Test d'Anderson-Darling

Ce test repose sur l'écart d'Anderson-Darling $A_n^2$ défini comme suit :

$$A_n^2 = n \int_0^\infty \frac{(F_n^*(x) - F(x))^2}{F(x)(1 - F(x))} dF_0(x). \qquad (33)$$

Cet écart permet de tester (32), l'expression opérationnelle de $A_n^2$ étant :

$$A_n^2 = -n - \frac{1}{n} \sum_{i=1}^{n} (2i-1)\left\{\ln F(x_{(i)}) + \ln\left[1 - F(x_{(n-i+1)})\right]\right\} \qquad (34)$$

où $x_{(i)}$ désigne la $i$-ème plus petite réalisation de $X$ dans l'échantillon.

On rejettera $H_0$ si

$$-n - \frac{1}{n}\sum_{i=1}^{n}(2i-1)\left\{\ln F_0(x_{(i)}) + \ln\left[1 - F_0(x_{(n-i+1)})\right]\right\} \qquad (35)$$

est supérieur à une valeur que la variable aléatoire $A_n^2$ a une probabilité α de dépasser.

## ACKNOWLEDGEMENTS

Les auteurs tiennent à remercier Pierre Devolder pour ses remarques qui nous ont aidées dans la finalisation de cet article.